\RequirePackage{fix-cm}

\documentclass[twocolumn,epjc3]{svjour3}

\smartqed  
\RequirePackage{fix-cm}

\smartqed  
\RequirePackage{graphicx}
\usepackage{graphicx,amsmath,amssymb,xcolor}

\journalname{}
\usepackage{url}
\newcommand{\beq}{\begin{equation}}
	\newcommand{\eeq}{\end{equation}}
\newcommand{\bea}{\begin{eqnarray}}
	\newcommand{\eea}{\end{eqnarray}}

\newcommand{\ce}{{\cal{E}}} \newcommand{\cl}{{\cal{L}}}
\newcommand{\cb}{{\cal{B}}}

  \def\cp{\pi}

\def\d{\dif}

\def\BB{{\cal B}}

\def\Uf{{\cal U}}

\def\tt{\theta}

\def\cp{P}

\providecommand{\dif}{\mathrm{d}} \def\d{\dif}

\def\aprx{\sim}
\allowdisplaybreaks
\graphicspath{{pic/}}

\begin{document}

\title{Charged particle motion and acceleration around Kerr-MOG black hole}

\author{
Saeed Ullah Khan\thanksref{e1,addr1,addr2} \and
Javlon Rayimbaev \thanksref{e2,addr21,addr22,addr23,addr24} \and Zden\v ek Stuchl\'ik \thanksref{e4,addr4} 
}

\thankstext{e1}{Corresponding author e-mail: saeedkhan.u@gmail.com}
\thankstext{e2}{e-mail: javlonrayimbaev6@gmail.com}
\thankstext{e4}{zdenek.stuchlik@fpf.slu.cz}

\institute{
{College of Mathematics and Statistics, Shenzhen University, Shenzhen 518060, China\label{addr1}}
\and
{College of Physics and Optoelectronic Engineering, Shenzhen University, Shenzhen 518060, China\label{addr2} \and School of Mathematics and Natural Sciences, New Uzbekistan University, Movarounnahr street 1, Tashkent 100000, Uzbekistan \label{addr21} \and School of Engineering, Central Asian University, Tashkent 111221, Uzbekistan  \label{addr22} \and Faculty of Computer Engineering, Tashkent University of Applied Sciences, Gavhar Str. 1, Tashkent 100149, Uzbekistan \label{addr23} \and Institute of Fundamental and Applied Research, National Research University TIIAME, Kori Niyoziy 39, Tashkent 100000, Uzbekistan \label{addr24}, \and Research Centre for Theoretical Physics and Astrophysics, Institute of Physics, Silesian University in Opava, Bezru\v covo n\' am. 13, CZ-74601 Opava, Czech Republic \label{addr4} }
 }

\date{Received: date / Accepted: date}
\maketitle

\begin{abstract}
One of the most important issues in relativistic astrophysics is to explain the origin mechanisms of (ultra)high energy charged particle components of cosmic rays. Black holes (BHs) being huge reservoirs of (gravitational) energy can be candidates for such particle sources. The main idea of this work is to study the effects of scalar-tensor-vector gravity (STVG) on particle acceleration by examining charged particle dynamics and their acceleration through the magnetic Penrose process (MPP) near magnetized Kerr-MOG BHs. First, we study the horizon structure of the BH. Also, we study the effective potential to gain insight into the stability of circular orbits. Our results show that the magnetic field can extend the region of stable circular orbits, whereas the STVG parameter reduces the {instability} of the circular orbit. The motion of charged particles around the magnetized BH reveals various feasible regimes of the ionized Keplerian disk behavior. 
Thus, from the examination of particle trajectories we observe that at fixed values of other parameters, the Schwarzschild BH captures the test particle; in the case of Kerr BH, the test particle escapes to infinity or is captured by the BH, while in Kerr-MOG BH, the test particle is trapped in some region around BH and starts orbiting it.
On investigating the MPP, we found that with increasing magnetic field, the behavior of orbits becomes more chaotic. As a result, the particle escapes to infinity more quickly.
\end{abstract}



\tableofcontents

\sloppy
\section{Introduction}\label{sec1}
In recent years, the process of exploring unseen components of our Cosmos got considerable interest from numerous scientists. The investigation of different cosmic data sets showed that dark energy (DE) plays a dominant role in the accelerating growth of our Cosmos \cite{Peebles:2002gy}. Theories and astrophysical observations related to cosmology and the early universe indicate that over $70\%$ of the whole universe is made up of DE \cite{Caldwell:2009zzb}, which can be described by a quintessential field or by a repulsive cosmological parameter $\Lambda>0$ \cite{Ostriker:1995su,Caldwell:1997,Faraoni:2000wk,Armendariz-Picon:2000nqq,Faraoni2000PhRvD,Adami:2012uv}. Besides, there exists another mysterious object termed dark matter (DM), which is supposed to be comprised of at least $20\%$ matter content of the Universe. Such matter consists of only weakly interacting particles and, as a result, does not interact with light and electromagnetic forces. Various experimental research has been carried out for the direct investigation of DM, but till now, there has been no appreciative outcome obtained \cite{CoGeNT:2010ols}.
Hence, modified theories of gravity can be used to investigate such a puzzling character of the DM. These theories are derived from Einstein's general theory of gravity \cite{Capozziello:2011et}.
	
Milgrom \cite{Milgrom:1983ca} was among the pioneers who modified Newtonian gravity to Modified Newtonian Dynamics (MOND). The MOND model may be used to explore the rotating curves of galaxies, but it may not be implemented for the gas density and temperature profiles of galaxy clusters without DM \cite{Angus:2007mn}. An extended version of the MOND in Tensor-vector-scalar gravity (TeVeS) is assumed to be a possible replacement for General Relativity without DM \cite{bekenstein2004relativistic}. In the case of a nonrelativistic weak acceleration limit, the above-mentioned model gives rise to the MOND, whereas its nonrelativistic strongly accelerating framework is Newtonian.

An alternative DM model, the scalar-tensor-vector gravity (STVG), was introduced in \cite{moffat2006scalar}. Along with the matter action and Einstein-Hilbert term, the field of three massive vectors and scalars are added to the action. The said theory can be used to analyze the solar system, gravitational lensing, and the rotational curves of galaxies, as well as the dynamics of galaxy clusters in the absence of DM \cite{MOFFAT2007,Moffat:2006rq}. By investigating MOG BHs, Moffat \cite{Moffat:2015kva} deduced that the parameter $\alpha$ increases the shadow radius. Lee and Han \cite{Lee:2017fbq}, by studying the Kerr-MOG BHs, found that $\alpha$ contributes to the innermost stable circular orbit's (ISCO) radius. Recently, Khan and Ren looked at the shadow cast by BHs in the existence of a cosmological constant $\Lambda$ and the quintessential DE extrapolate that both $\gamma$ and $\Lambda$ contribute to the shadow radius \cite{Khan:2020ngg,Khan:2020hdq}.

The weak gravitational lensing properties of Schwarzschild-MOG BHs in the presence of plasma medium are studied in \cite{Atamurotov2021EPJC}. The dynamics of test charged \& magnetized particles near regular \& Schwarzschild BHs and magnetic fields in MOG have been explored in Refs.\cite{Rayimbaev2023EPJP,Haydarov2020EPJC,Murodov2023arXiv231008046M,Turimov2023PhLB..84338040T,2023MPLA...3850071B}. S2 star testes \cite{Turimov2022MNRAS.516..434T} and twin QPOs around regular BHs \cite{Rayimbaev2021Galax} and vacuum and plasma magnetosphere of rotating magnetized neutron stats and deathline conditions of radio-loud pulsars have also been investigated in MOG \cite{Rayimbaev2020PhRvD.102b4019R}.

The presence of a magnetic field around BHs has essential consequences on the process of accretion, as well as charged matter. It is found that BHs possess an accretion disk generated by conducting plasma, and their motion could constitute a regular magnetic field. Recent findings show the existence of strong magnetic fields near a supermassive in the Galaxy center that has no association with the accretion disk \cite{eatough2013strong}. Hence, BHs could be immersed in an external magnetic field having a composite formation near a BH's horizon. At the same time, its nature will be simple and adjacent to a magnetic field of homogeneous nature at a larger finite distance \cite{Martin-2015}. Kovar et al. \cite{Kovar-2014} observed that BH in the equatorial plane of a magnetar could be submerged in a homogeneous magnetic field if the magnetar is placed at a distant location. Therefore, in the current work, our aim is to focus on BH immersed in a uniform external magnetic field, also called the Wald solution of the magnetized BH \cite{Wald}.

The dynamics of both electrically charged and neutral test particles near BHs are of ample interest in the field of relativistic astrophysics. In particular, charged-particle dynamics are much more interesting, because they play a crucial role in understanding the consequences of the magnetic field on the accretion process. The equations of electrically charged particle dynamics outlined by the Reissner-Nordstr\"{o}m or Kerr-Newman spacetime can easily be separated and integrated \cite{Carter-1973}. Such dynamical motion of particles has been studied in various articles \cite{Bicak,Stuchlik-1999,Stuchlik-2009,Pugliese-2011,Pugliese-2013,2021EPJC...81..269N}  Konoplya \cite{Konoplya:2006qr}, by examining particle dynamics around a magnetized BH, and deduced that the tidal charge has a strong influence on the motion of masses as well as the massless motion of particles. An abundant and detailed investigation of the charged particle motion in magnetized BHs can be found in \cite{Preti-2004,Kopacek-2010,Zahrani-2013,Shiose-2014,Panis-2019,Stu-etal:2020:Universe:}. Particle collision in the ergosphere and the dynamics of particles are examined in Kerr and Kerr-Newman-Kasuya BHs, respectively \cite{Khan1,Khan2,Khan3}. Recently, Khan and Chen, by studying charged particle dynamics around Kerr BH in a split monopole magnetic field, investigated the position of stable circular orbits and deduced that the positive magnetic field increases the stability of effective potential \cite{Khan:2023ttq}
	
The weak magnetic field has negligible effects on the motion of neutral particles or the background geometry. However, the magnetic field has a notable impact on the trajectory of a charged test particle. Thus, for charged test particles having charge $q$ and mass $m$ revolving near a BH of mass $M$ that is enveloped in a magnetic field of intensity $B$, a dimensionless quantity ${\cal B}=qB GM/mc^4$ may be defined as the relative Lorenz force \cite{Frolov,Tursunov-KOrbit}. This quantity could also have a significant value in an inadequate magnetic field, owing to the greater amplitude of a particular charge $q/m$. We consider that a "charged particle" may refer to any material ranging from electrons to some sort of charge inhomogeneity revolving in an accretion disk's innermost location \cite{Kolos1}.

One of the most realistic energy-release processes from rotating BHs is the classical Penrose process, which was suggested by R. Penrose in 1969 \cite{Penrose69}. The main idea is that a neutral particle decays by two neutral parts at the ergoregion of a rotating Kerr BH and one of them falls down the BH center with negative energy, while the other one leaves the region with positive energy and higher than the initial ("mother") particle's energy. In recent years, several approaches and modifications of Penrose processes have been developed in gravity theories. For example, the magnetic and electric Penrose processes have been suggested in Refs. \cite{Dadhich18,wagh85,2021arXiv210805116D,2018PhLB..778...54M,2023EPJC...83..506K,2021PhRvD.104h4099T,2021Univ....7..416S}, considering the decayed particles are spinning and charged. 

Our primary objective in this work is to examine charged-particle dynamics and particle acceleration near a Kerr-MOG BH drowned in a nearby homogeneous magnetic field. The essential outlines of our article are as follows. In section \ref{sec2}, we briefly review the Kerr-MOG BH spacetime and its horizons. Section \ref{sec3} investigates charged particle motions around a Kerr-MOG BH. The fourth section \ref{sec4} of our article explores particle acceleration in the MPP. In the last section \ref{Concluding}, we conclude our study with closing comments.

Throughout this paper, we use the geometrical unit system and take $G_N=c=1$.
\section{Kerr-MOG black hole}\label{sec2}

\subsection{STVG field}

{The gravitational field action in the STVG theory includes GR $S_{G}$, matter (pressure-less) $S_{M}$, vector field $S_{\Phi}$ and scalar field $S_{S}$ terms \cite{Mureika16}:
\begin{equation}
    \label{action}
 S = S_{G}+S_{\phi}+S_{S}+S_{M}\ ,
\end{equation}
where 
\begin{eqnarray}
&& S_{G} = \frac{1}{16\pi}\int \frac{1}{G} (R+2 \Lambda) \sqrt{-g}d^4x,
\\ && S_{\Phi} = -\frac{1}{4\pi} \int \left[{\cal K} + {\rm V}(\Phi)\right] \sqrt{-g}d^4x\ , 
\\
\nonumber
&& S_{S} = \int \frac{1}{G} \Bigg[\frac{1}{2} g^{\alpha \beta} \left( \frac{\nabla_{\alpha}G \nabla_{\beta}G}{G^2} + \frac{\nabla_{\alpha}\mu \nabla_{\beta}\mu}{\mu^2} \right)
\\ &&- \frac{V_{G}(G)}{G^2} -\frac{V_{\mu}(\mu)}{\mu^2} \Bigg] \sqrt{-g}d^4 x\ ,
\\
&&  S_{M} = -\int (\rho \sqrt{u^{\mu}u_{\mu}}+{\cal Q}u^{\mu}\Phi_{\mu})\sqrt{-g}d^4 x + J^{\mu} \Phi_{\mu}\ ,
\end{eqnarray}
with $R = g^{\mu\nu} R_{\mu\nu}$ is the Ricci scalar, $\Lambda$ is the cosmological constant, $g\equiv$det$(g_{\mu\nu})$ is the determinant of the metric tensor, $\nabla_{\alpha}$ is the covariant derivation, ${\cal K}$ is the kinetic term for the scalar field $\Phi_{\mu}$ which reads as 
\begin{equation}
    {\cal K} = \frac{1}{4}B^{\mu\nu} B_{\mu\nu}\ ,
\end{equation}
where { $B^{\mu\nu} = \partial _{\mu} \Phi_{\nu} -\partial _{\nu} \Phi_{\mu}.$} The covariant current density is defined to be 
\begin{equation}
 J^{\mu} =\kappa T_M^{\mu\nu} u_{\nu}\ ,
 \label{current-dens}
\end{equation}
where $T_M^{\mu\nu}$ is the energy-momentum tensor for matter with $\kappa = \sqrt{\alpha G_N},$ $\alpha = (G-G_N) /G_N$ is a parameter defining the scalar field, $G_N$ is Newtonian gravitational constant, $u^{\mu} = dx^{\mu} /d\tau$ is a timelike velocity and $\tau$ is the proper time a long time like geodesic. The perfect fluid energy-momentum tensor for matter is given by 
\begin{equation}
 T^{M\mu\nu} = (\rho_M + p_M) u^{\mu}u^{\nu} -p_Mg^{\mu\nu}\ ,
 \label{tensor-energy}
\end{equation}
where $\rho_M$ and $p_M$ are the density and pressure of matter, respectively. 
From Eqs. (\ref{current-dens}) and (\ref{tensor-energy}) using { $u_\mu u^\mu=-1$,} we get 
\begin{equation}
 J^{\mu}={k} \rho_M u^{\mu}\ . 
\end{equation}
For the matter-free and pressureless MOG field ($T_M^{\mu\nu}=0$) in the asymptotically flat (zero-cosmological constant) spacetime, the field equation takes the form 
\begin{equation}
    G_{\mu\nu}=-\frac{8\pi G}{c^4}T^\Phi_{\mu\nu}\ ,
\end{equation}
where $T^\Phi_{\mu\nu}$ is the tensor of massive-vector field. The observational data from galaxy and cluster dynamics show that the mass of the particles of the field $\Phi$ is about $m_\Phi=2.6\times10^{-28}$ eV, and it is almost zero \cite{Moffat13}. One may assume that the vector field is an analogue of the electromagnetic field and its field tensor is defined as
\begin{equation}
    T^\phi_{\mu\nu}=-\frac{1}{4\pi}(B_\mu^\alpha B_{\nu \alpha} - \frac{1}{4} g_{\mu \nu} B^{\alpha\beta}B_{\alpha\beta})
\end{equation}
with 
{
\begin{eqnarray}
 && \Delta_\mu B^{\mu\nu}=0\ , \\ && \Delta_\alpha B^{\mu\nu}+\Delta_\nu  B^{\mu\alpha}+\Delta_\mu B^{\alpha\nu}=0\ .
\end{eqnarray}}
The above assumptions imply that the potential term of the action $S_\Phi$ is zero (${\rm V}(\phi)=(1/2)\mu \Phi_\mu\Phi^\mu=0$), so it has only kinetic term, and one may consider the kinetic term is a function of the massive-vector field invariant ${\cal B}=B_{\mu\nu}B^{\mu\nu}$ as ${\cal K}=f({\cal B})$. 

\subsection{Kerr-MOG BH solution}

This section aims to study Kerr-MOG BH submerged in an externally uniform magnetic field. The MOG field equations are axially symmetric, stationary, and asymptotically flat. In Boyer-Linquist coordinates, its spacetime geometry can be expressed as \cite{Moffat:2014aja}
\beq
\d s^2 = g_{tt}\d{t}^2 + g_{rr}\d{r}^2 + g_{\theta\theta}\d\theta^2 + g_{\phi\phi}\d\phi^2 +2g_{t\phi}\d{t}\d\phi, \label{Kerr-MOGMetric}
\eeq
with
\bea
&&g_{tt} = -\left(\frac{\Delta-a^{2}\sin^{2}\theta}{\rho^2}\right), \quad
g_{rr} = \frac{\rho^2}{\Delta}, \quad g_{\theta\theta} = \rho^2,\nonumber\\
&&g_{\phi\phi} = \frac{\sin^{2}\theta}{\rho^2}\left[(a^{2}+r^{2})^{2}-a^{2}\sin^{2}\theta \Delta\right], \nonumber\\
&&g_{t\phi}=\frac{a\sin^{2}\theta}{\rho^2}\left[\Delta-(a^{2}+r^{2})\right] ,
\label{MetricCoef}
\eea
in which
\bea
&&\Delta = r^2 - 2GMr + a^2 + \alpha G_N G M^2, \nonumber\\
&&\rho^2 = r^2 + a^2 \cos^2\theta.\label{Delta}
\eea
In the above expressions, $G=G_N(1+\alpha)$ and $G_N$, respectively, define the constant of gravity and Newton's constant of gravity. The parameter $a$ represents BH's spin, while the dimensionless parameter $\alpha$ describes the gravitational field's strength. Moreover, $M$ represents BH's mass and could be related to the Arnowitt–Deser–Misner (ADM) mass by the relation $\mathcal{M}_{\alpha}=(1+\alpha)M$ \cite{Sheoran:2017dwb}.
By incorporating the value of $G$ and utilizing the dimensionless quantities $r\rightarrow r/G_{N}M$, $a\rightarrow a/G_{N}M$ and $\alpha \rightarrow \alpha/G_{N}M$ in Eq. \eqref{Delta}, the value of $\Delta$ takes the form 
\beq\label{D1}
\Delta=r^{2}-2(1+\alpha)r+a^{2}+\alpha(1+\alpha). 
\eeq 

While in the rotating spacetime, the vector potential can be extended as
\begin{eqnarray}\label{vpot}
    \Phi_\mu=\frac{\sqrt{\alpha} Mr}{\Sigma}(-1,0,0,a \sin^2{\theta})
\end{eqnarray}
\subsection{Event horizon properties}
The spacetime metric (\ref{Kerr-MOGMetric}), diminishes to the Kerr spacetime on letting $\alpha=0$; to the Schwarzschild-MOG BH on setting $a=0$; to the Schwarzschild BH for $\alpha=a=0$. 

From the observations of the Solar system, the restriction on parameter $\alpha$ can be read as \cite{moffat2006scalar,LopezArmengol:2016irf}
\beq
\alpha_\odot < 1.5 \times 10^5 \textit{cm} \times \frac{c^2}{G_N M_\odot} \approx 1 \label{Restriction}
\eeq

\begin{figure*}
\includegraphics[width=0.44\textwidth]{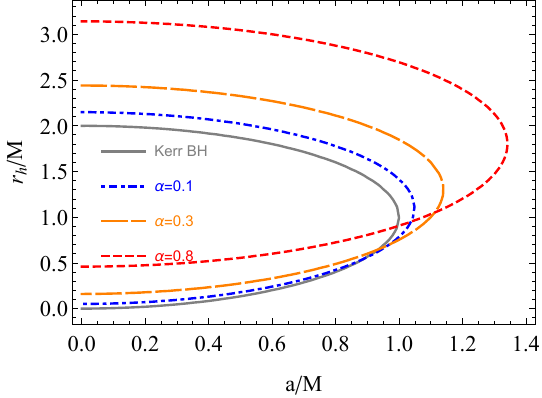}
\includegraphics[width=0.47\textwidth]{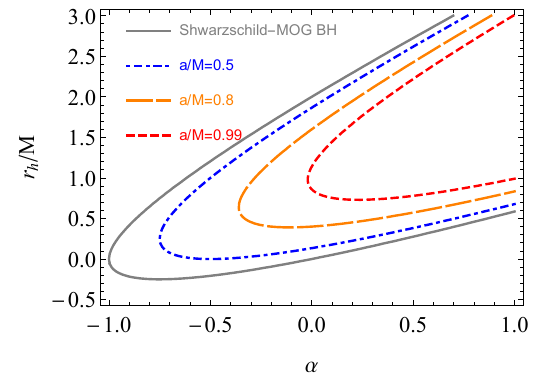}
\caption{Graphical description of the BH horizons at various combinations of spin parameter $a$ left and $\alpha$ right panel.}\label{Hor}
\end{figure*}
Figure \ref{Hor} describes the geometrical structure of BH horizons vs. the spin parameter $a$ on the left side, whereas vs. the parameter $\alpha$ on the right side. The graphical behavior shows that Kerr BH has the smallest horizon, whereas Schwarzschild-MOG BH has the largest horizon. The graphs also depict that $\alpha$ results in increasing BH's horizons, whereas $a$ results in shrinking the horizons.

One can obtain possible values of MOG and spin parameters that can provide a horizon in the spacetime of the Kerr-MOG BH by setting $\Delta=0$ and $\partial_r\Delta=0$.
\begin{figure}
\includegraphics[width=0.5\textwidth]{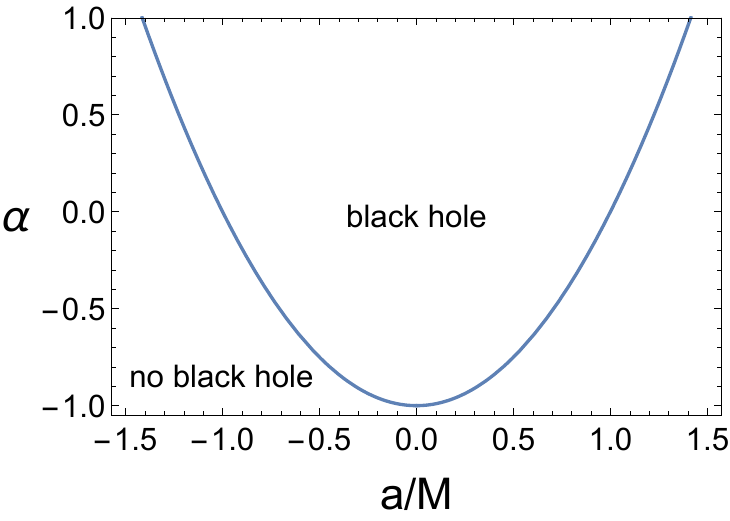}
\caption{BH vs no BH regions in $\alpha-a$ space.}\label{alphavsa}
\end{figure}

In Fig.\ref{alphavsa}, we demonstrate a set of values of $\alpha$ and $a$ that correspond to BH (BH with horizon) and no BH (an object without no-horizon). The blue line implies extreme values for the BH spin correspond to the maximum value of the MOG field parameter. It is observed that the extreme values of $a_{extr}$ increase with the increase $\alpha_{max}$ quadratic form. In our further analyses, we use these results in choosing test values for spin and MOG parameters: above, the blue line corresponds to BH solutions. 
\subsection{Magnetization of Kerr-MOG black holes}
We assume that the Kerr MOG BH is immersed in an external asymptotically uniform magnetic field and the magnetic field $B$ is orthogonal to the equatorial plane being oriented along the $z$ axis. It is worse to note that the magnetic field could be expressed with the help of the electromagnetic 4-vector potential $\mathcal{A}_\mu$ using Wald's approach. Henceforth, its non-vanishing terms can be described as \cite{Frolov,Stuchlik2020Univ....6...26S}
\bea
\mathcal{A}_t&=&\frac{B}{2}(g_{t\phi}+2ag_{tt})-\frac{\mathcal{Q}}{2}g_{tt}-\frac{\mathcal{Q}}{2},\\
\mathcal{A}_{\phi}&=&\frac{B}{2}(g_{\phi\phi}+2ag_{t\phi})-\frac{\mathcal{Q}}{2}g_{t\phi}.
\eea
The induced electrical charge $\mathcal{Q}$ is called Wald charge and explains the induced electric potential difference between the BH horizon and infinity due to the presence of the external magnetic field and rotation of the spacetime. The maximum value induced BH charge produced due to the BH spin grabs the value of Wald $\mathcal{Q}_W=2aBM$ \cite{Wald} In principle, BHs with a maximum Wald charge  possess the reduced electromagnetic potential as given below
\bea\label{Elec-MagPotential}
\mathcal{A}_t=\frac{B}{2}g_{t\phi}-\frac{\mathcal{Q}_W}{2},\quad
\mathcal{A}_{\phi}=\frac{B}{2}g_{\phi\phi}.
\eea
\section{Geodesics around Kerr-MOG spacetime}\label{sec3}
This section aims to explore the characteristics of the photon region around a Kerr-MOG BH, with the assumptions of $p^{\mu}p_{\mu}=-m^2$. Therefore, by applying the technique of separation of variables, the corresponding geodesic motion of the metric \eqref{Kerr-MOGMetric} could be expressed by the Hamilton–Jacobi equation as

\begin{eqnarray}\nonumber
\mathcal{H} &=& -\frac{\partial S}{\partial \tau}\\ \nonumber\label{Hamiltonian} 
  &=& \frac{1}{2}g_{\mu\nu}(P_\mu-q\mathcal{A}_\mu+\tilde{q}\Phi_\mu)(P_\nu-q\mathcal{A}_\nu+\tilde{q}\Phi_\nu)\\ &+&\frac{1}{2}m^2.
\end{eqnarray}	

Where $\tilde{q}=\sqrt{\alpha}m$ is gravitational test particle charge.
In Eq. \eqref{Hamiltonian}, $\mathcal{H}$ and $S$, respectively, denote the Hamiltonian and Jacobi actions. Moreover, $m^2$ is the rest mass, while $p_\mu={\partial S}/{\partial{x}^{\mu}}$, is the four-momentum of particles.
Due to spacetime symmetry, the preserved energy and angular momentum may be described in the form of:
\begin{eqnarray}\label{Energy}
-\mathcal{E}&=&p_{t}=g_{tt}\dot{t}+g_{t\phi} \dot{\phi}+q\mathcal{A}_t+\Tilde{q}\Phi_t,\\ \label{Momentum}
\mathcal{L}&=&p_{\phi}=g_{t\phi}\dot{t}+g_{\phi\phi}\dot{\phi}+q\mathcal{A}_\phi+\Tilde{q}\Phi_\phi.
\end{eqnarray}
The dot denotes the derivative regarding the proper time $\tau$. Therefore, by using $\mathcal{E}=E/m$, $\mathcal{L}=L/m$ and a particular charge $\bar{q}=q/m$ the Hamiltonian in Eq. \eqref{Hamiltonian} modifies to
\beq
\mathcal{H}=\mathcal{H}_p(r,\theta)+\frac{1}{2}g^{rr}p_r^2+\frac{1}{2}g^{\theta\theta}p_\theta^2,\label{M-Hamiltonian}
\eeq
with
\bea\nonumber
\mathcal{H}_p(r,\theta)&=&\frac{1}{2}\big[g^{tt}(\mathcal{E}+\bar{q} \mathcal{A}_t +\Tilde{q}\Phi_t)^2\\ \nonumber &+&g^{\phi\phi}(\mathcal{L}-\bar{q} \mathcal{A}_\phi -\Tilde{q}\Phi_\phi)^2\\\nonumber
&-&2g^{t\phi}(\mathcal{E}+\bar{q} \mathcal{A}_t +\Tilde{q}\Phi_t)(\mathcal{L}-\bar{q} \mathcal{A}_\phi -\Tilde{q}\Phi_\phi)+1\big].
\eea
In the above expression, $\mathcal{H}_p(r,\theta)$ represents the potential part of the Hamiltonian. Utilizing the Eqs. \eqref{Energy} and \eqref{Momentum}, the value of $\dot{t}$ and $\dot{\phi}$ can be obtained as
\bea
&&\dot{t}=-\frac{g_{t\phi}(\mathcal{L}-\bar{q} \mathcal{A}_\phi-\Tilde{q}\Phi_\phi)+g_{\phi\phi}(\mathcal{E}+\bar{q} \mathcal{A}_t +\Tilde{q}\Phi_t)}{g_{tt}g_{\phi\phi}-g_{t\phi}^2},\\
&&\dot{\phi}=\frac{g_{tt}(\mathcal{L}-\bar{q} \mathcal{A}_\phi-\Tilde{q}\Phi_\phi)+g_{t\phi}(\mathcal{E}+\bar{q} \mathcal{A}_t + \Tilde{q}\Phi_t)}{g_{tt}g_{\phi\phi}-g_{t\phi}^2}.
\eea
The charged test particle dynamics could be restricted with the energetic boundaries imposed by the $\mathcal{H}=0$ constraints. By making use of the energy condition, the effective potential may be associated with the specific energy for circular orbits ($\dot{r}=0$) as $\mathcal{E}=\Uf_{eff}(r, \theta)$ and in the Kerr-MOG BH simplifies to\cite{Kopa-Veff}
\beq \label{effpoteq}
\Uf_{eff}(r, \theta)=\frac{-\beta+\sqrt{\beta^2-4\delta \gamma}}{2\delta},
\eeq
here
\bea \nonumber
\delta &=& -g^{tt}, \\\nonumber
\beta & =& -2[g^{tt}(\bar{q}\mathcal{A}_t+\Tilde{q}\Phi_t)-g^{t\phi}(\mathcal{L}-\bar{q}\mathcal{A}_\phi-\Tilde{q}\Phi_\phi)], \\ \nonumber
\gamma &= &-g^{tt}(\bar{q}\mathcal{A}_t+\Tilde{q}\Phi_t)^2+2g^{t\phi}(\bar{q}\mathcal{A}_t +\Tilde{q}\Phi_t)\\ \nonumber
&& \times(\mathcal{L}-\bar{q}\mathcal{A}_\phi - \Tilde{q}\Phi_\phi) -g^{\phi\phi}(\mathcal{L}-\bar{q}\mathcal{A}_\phi-\Tilde{q}\Phi_\phi)^2-1 \,.
\eea
The presence of magnetic fields around BHs produces chaotic behavior in the charged particle dynamics, except in the equatorial plane. On the equatorial plane, the minima and maxima of $\Uf_{eff}$, respectively, indicate stable and unstable circular orbits. The circular orbits of Kerr-MOG BH can be distinguished into four different classes \cite{Tursunov-KOrbit}, namely the prograde anti-Larmor orbit (PALO) for $\mathcal{L}>0, \mathcal{B}>0$; retrograde Larmor orbit (RLO) for $\mathcal{L}<0, \mathcal{B}>0$; prograde Larmor orbit (PLO) for $\mathcal{L}>0, \mathcal{B}<0$ and retrograde anti-Larmor orbit (RALO) for $\mathcal{L}<0, \mathcal{B}<0$. Here, $\mathcal{L}>0$ and $\mathcal{L}<0$ denote the co-rotating and counter-rotating particles, respectively, while $\mathcal{B}>0$ and $\mathcal{B}<0$, respectively, define the repulsive and attractive Lorentz forces.
	
The dynamics of charged particles can be bounded by the boundaries of the effective potential
\beq
\mathcal{E}= \Uf_{eff}(r, \theta; \mathcal{B}, \mathcal{L}). \label{energetics}
\eeq
\begin{figure*}
	\includegraphics[width=0.32\textwidth]{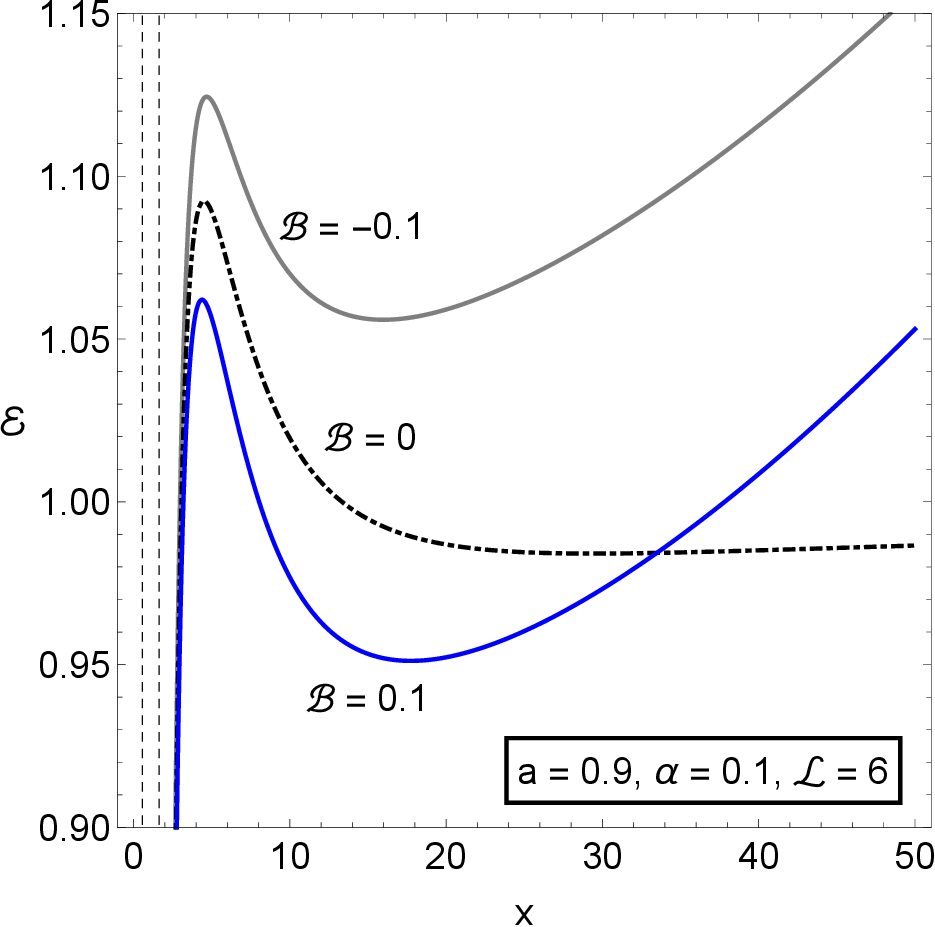}
	\includegraphics[width=0.32\textwidth]{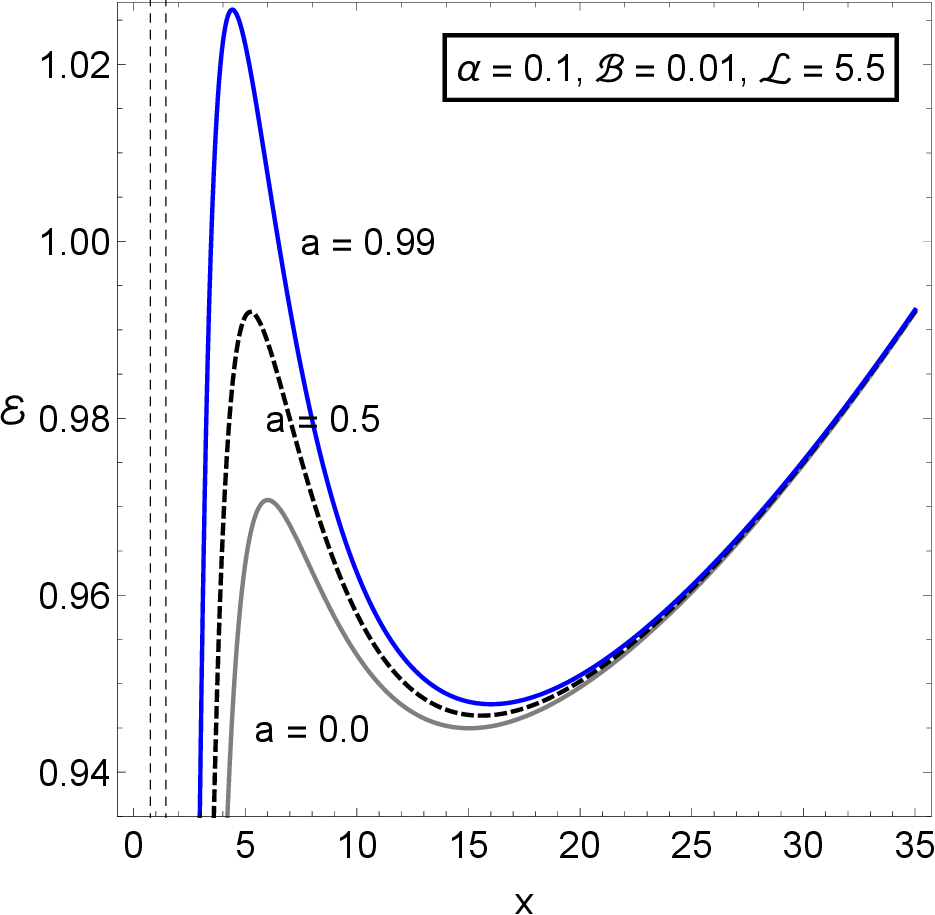}
	\includegraphics[width=0.32\textwidth]{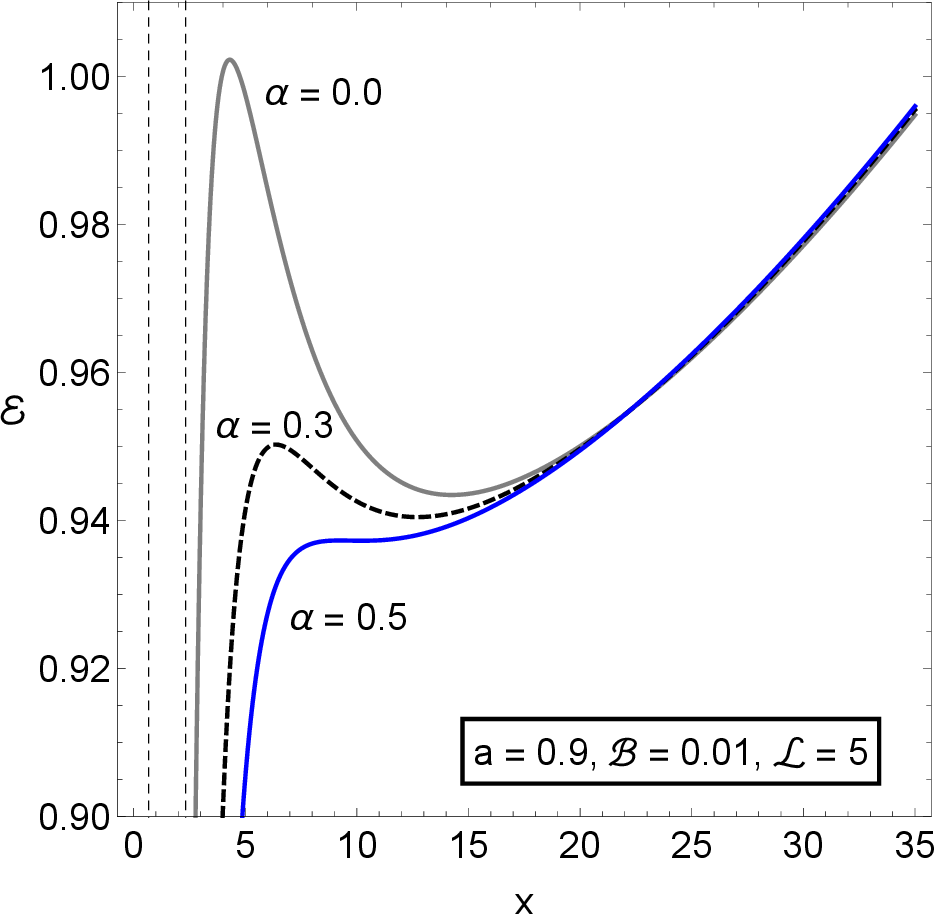}
	\caption{Graphical description of the effective potential at various discrete values along the radial coordinate x.}\label{veff}
\end{figure*}
{The behavior of the effective potential (\ref{effpoteq}) is investigated in Fig. \ref{veff}, which allows us to illustrate the general features associated with the motion of charged particles without explicitly computing the equation of motion \cite{Kolos:2015iva}. Also, the graphical behavior depicts that magnetic {field} stabilizes the circular orbits (see Fig. \ref{veff} first column). The particles coming from infinity with a higher spin parameter $a$ value required more energy to climb $\Uf_{eff}$ compared to a smaller value of $a$, while it is vice versa for the parameter $\alpha$ (for details, see the second and last column of Fig. \ref{veff}). The middle and right panels show that circular orbits are initially unstable but become stable along $\text{x}$ at a higher distance. It is also observed that the BH spin increases the instability, whereas $\alpha$ results in decreasing the instability near the BH horizon. Furthermore, the circular orbit's instability at the horizon in Kerr-MOG BH is higher than in Schwarzschild-MOG BH.}
\subsection{Charged particle trajectories}\label{sec3.1}
The Keplerian accretion disk could generate the trajectories of circular orbits near a BH while the ISCO determines its lower boundary. One can obtain the boundaries of particle dynamics by investigating the properties of the effective potential $\Uf_{eff}(r, \theta)$. Therefore, the circular orbits of the charged test particle could be acquired using stationary positions of the effective potential as
\beq
\partial_r \Uf_{eff}(r, \theta)=0, \quad
\partial_\theta \Uf_{eff}(r, \theta)=0.
\eeq
In the case of Schwarzschild BH, the local extrema of $\Uf_{eff}(r, \theta)$ are located only on the equatorial plane. On the contrary, in our case (Kerr-MOG BH), the local extrema in the equatorial plane can also determine the circular orbits \cite{Martin-2015}.
\begin{figure*}
	\begin{minipage}[b]{\textwidth}
		\includegraphics[width=\textwidth]{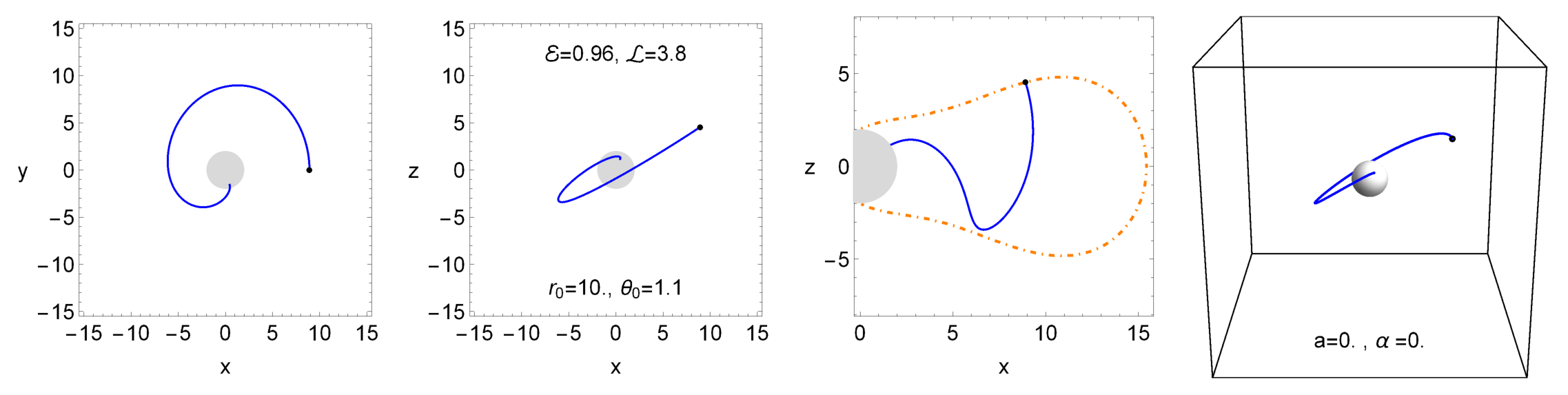}
	\end{minipage}\vspace{0.15cm}
	\begin{minipage}[b]{\textwidth}
		\includegraphics[width=\textwidth]{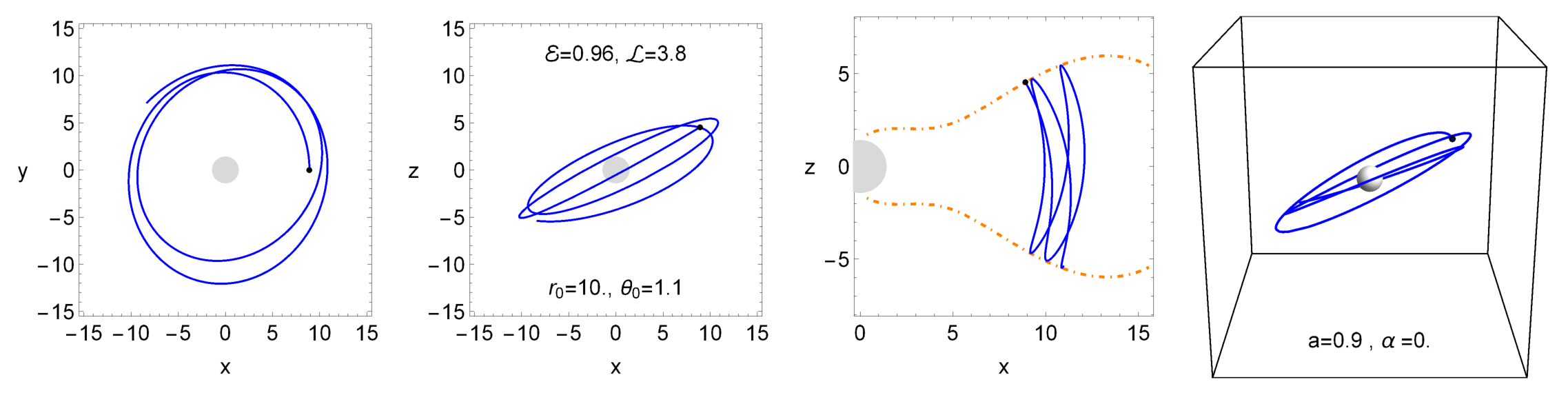}
	\end{minipage}\vspace{0.15cm}
	\begin{minipage}[b]{\textwidth}
		\includegraphics[width=\textwidth]{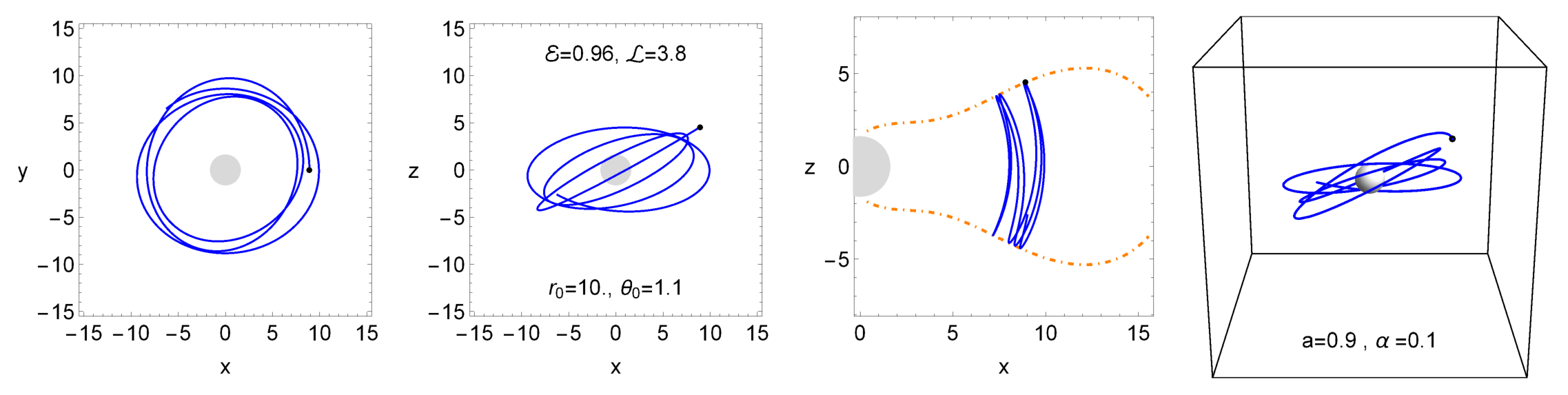}
	\end{minipage}
\caption{Test particle trajectories (solid curves) around BHs described by the shaded circle without considering external magnetic field ($\mathcal{B}$). Trajectories in the top, middle, and bottom rows indicate the Schwarzschild, Kerr, and Kerr-MOG BHs, respectively. The fourth column corresponds to the $3D$ particle trajectories, whereas the first and second columns describe its $2D$ orientations. Using the principles of energy and angular momentum conservation, we can describe the $4D$ configuration space ($t,x,y,z$) in a $2D$ graph (third column), in which the motion boundary is plotted via the effective potential (dotted-dashed curves).
	}\label{Traj1}
\end{figure*}
\begin{figure*}
	\begin{minipage}[b]{\textwidth}
		\includegraphics[width=\textwidth]{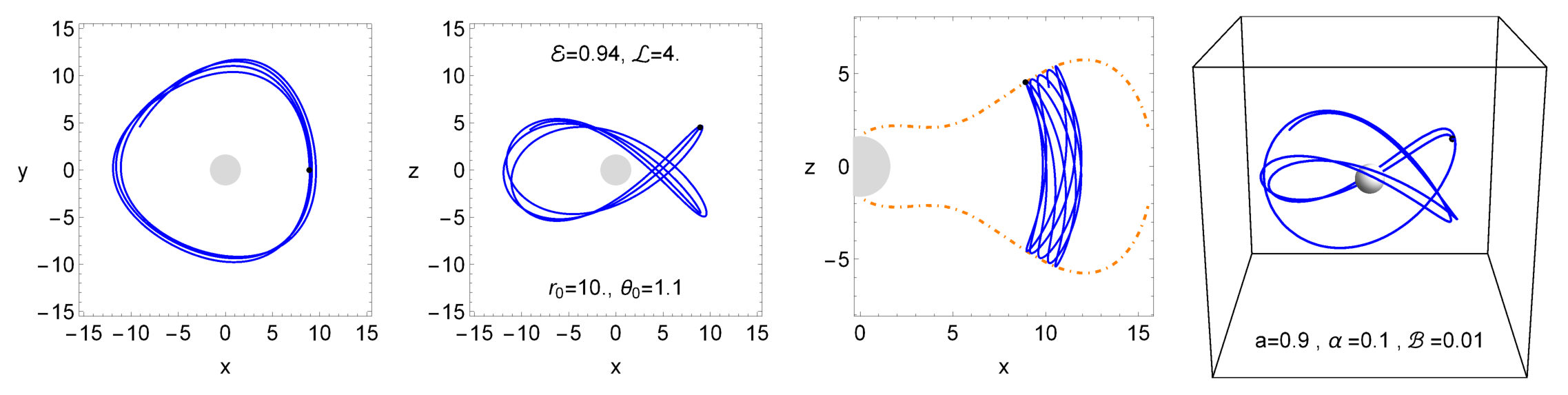}
	\end{minipage}\vspace{0.15cm}
	\begin{minipage}[b]{\textwidth}
		\includegraphics[width=\textwidth]{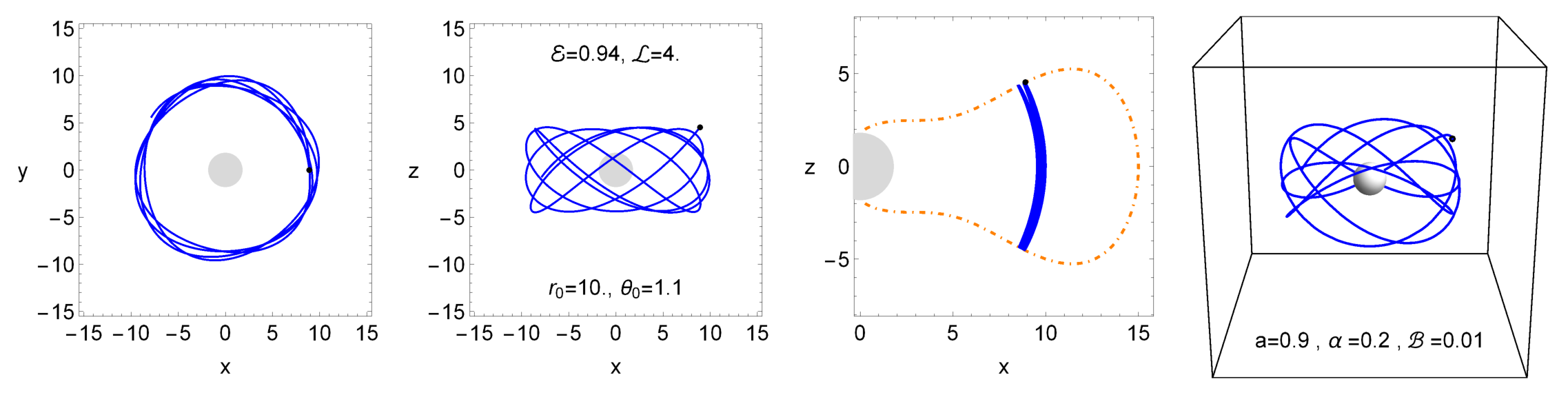}
	\end{minipage}\vspace{0.15cm}
	\begin{minipage}[b]{\textwidth}
		\includegraphics[width=\textwidth]{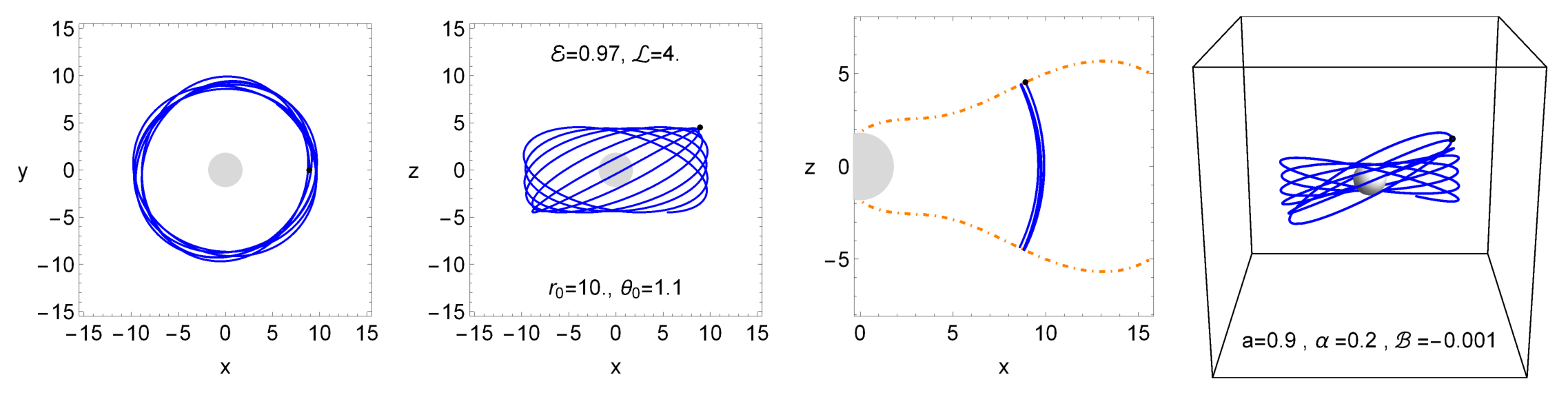}
	\end{minipage}\vspace{0.15cm}
	\begin{minipage}[b]{\textwidth}
		\includegraphics[width=\textwidth]{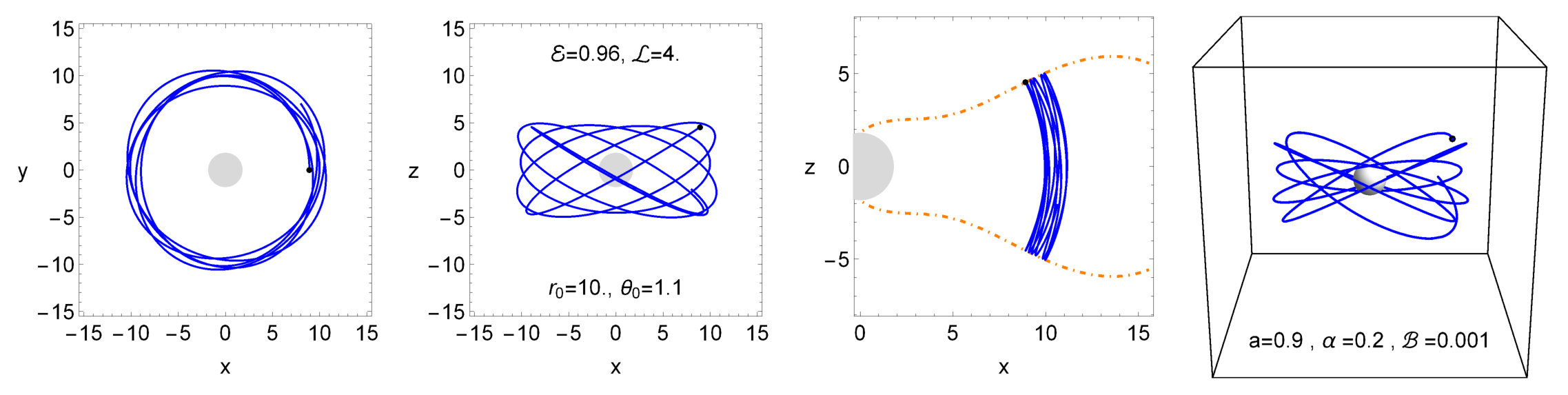}
	\end{minipage}
	\caption{Charged particle trajectories around Kerr-MOG BH with an external magnetic field at various discrete values of $\alpha$ (for details description see Fig. \ref{Traj1}).}\label{Traj2}
\end{figure*}
\begin{figure*}
	\begin{minipage}[b]{\textwidth}
		\includegraphics[width=\textwidth]{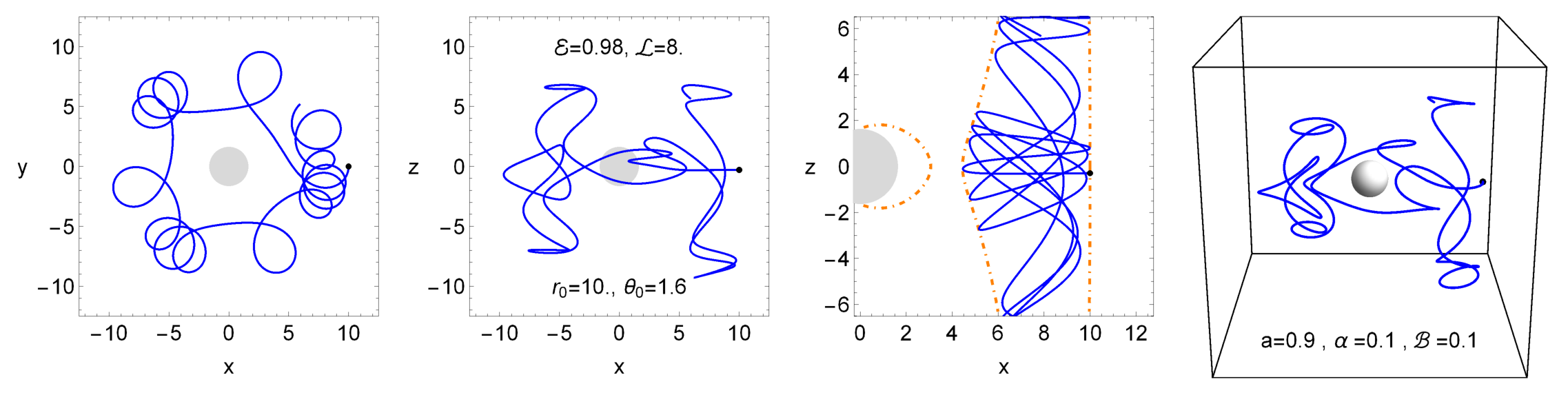}
	\end{minipage}\vspace{0.15cm}
	\begin{minipage}[b]{\textwidth}
		\includegraphics[width=\textwidth]{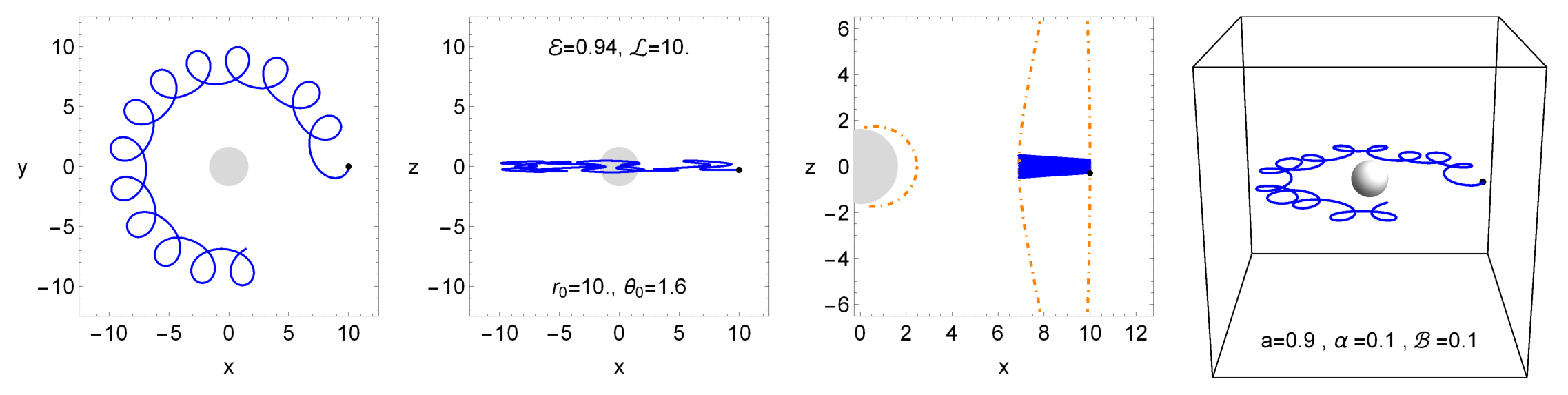}
	\end{minipage}\vspace{0.15cm}
	\begin{minipage}[b]{\textwidth}
		\includegraphics[width=\textwidth]{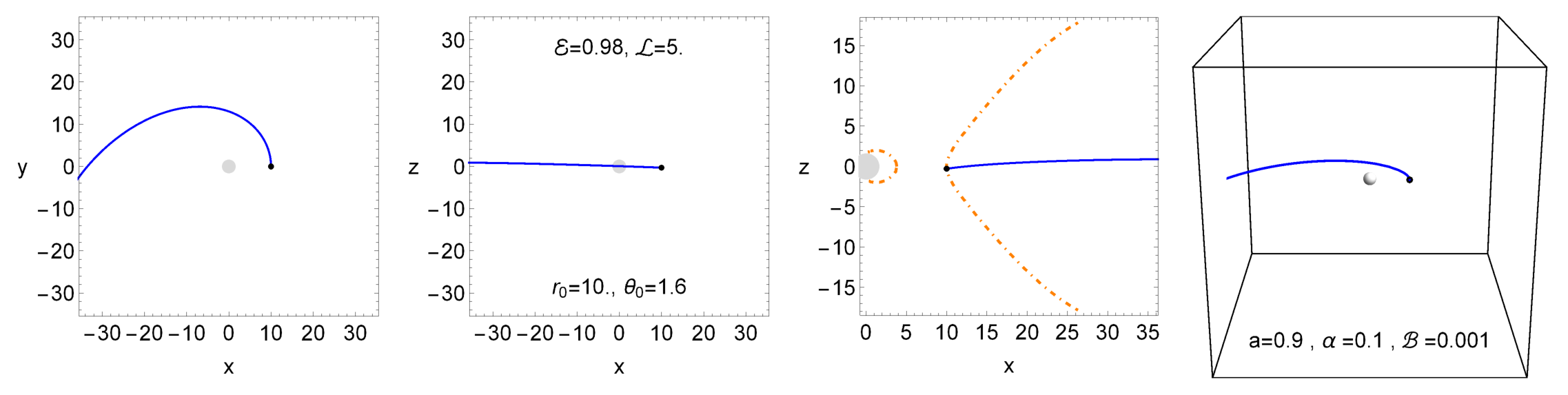}
	\end{minipage}
	\begin{minipage}[b]{\textwidth}
		\includegraphics[width=\textwidth]{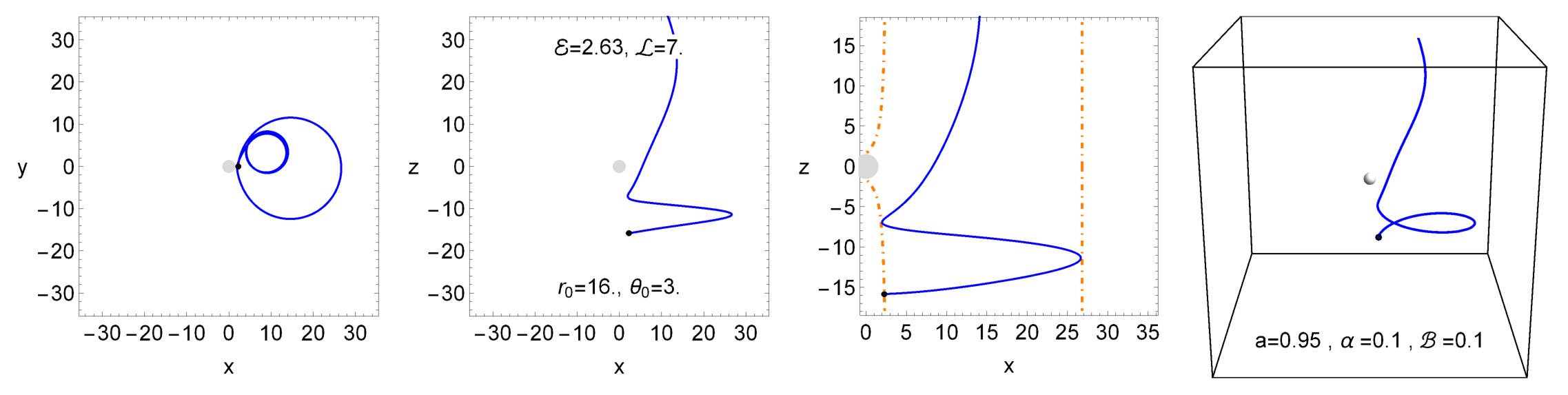}
	\end{minipage}
	\caption{ Charged test particle trajectories (solid curves) around a Kerr-MOG BH with an external magnetic field (for details description see Fig. \ref{Traj1}).}\label{Traj3}
\end{figure*}

In principle, charged particle motion near a BH immersed in uniform external magnetic fields has chaotic behavior. Although charged particle trajectories near stable circular orbits still possess a structured nature \cite{Kopacek-2010}. It has been observed that the charged particles also have structural trajectories on the equatorial plane, while their nature becomes chaotic as the angle of inclination of the beginning point (since the inclination of orbits changes in each period) varies from the equatorial plane.

The characteristic of effective potential may provide various kinds of energetic boundaries \eqref{energetics}. The graphical behavior of the test charged particle trajectories in the background of Schwarzschild, Kerr, and Kerr-MOG BHs are shown, respectively, in the top, middle, and bottom rows of Fig. \ref{Traj1} in the absence of the external magnetic field, at $\mathcal{B}=0$. 
From Fig \ref{Traj1} it can be observed that at the same value of other parameters, the Schwarzschild BH captures the test particle. In contrast, in the case of Kerr and Kerr-MOG, the test particle is trapped in some region around BH and starts orbiting the BH in circular orbits due to a decrease of gravitational forces in the spacetime and an increase of centrifugal forces. It means that in Kerr and Kerr MOG BH cases the falling rate of accreting matter is smaller than in the case of the Schwarzschild BH. {Also, the comparisons of the middle and bottom rows show that the presence of the MOG parameter makes chaotic (unstable) the trajectories (bounded orbits) and reduces the wideness between turning planes due to the fifth (MOG field) interaction.}

{In Figs. \ref{Traj2} and \ref{Traj3}, we have plotted the charged particle trajectories in the close environment of Kerr-MOG BH in the presence of an external asymptotically uniform magnetic field (with $\mathcal{B}\neq 0 $). We can observe various types of orbits with corresponding boundary conditions. The first one is related to the existence of an outer boundary, where the charged particle should be captured by the central magnetized BH. The second type is related to the existence of an inner boundary where the particle must escape to infinity. The third type is related to the existence of inner and outer boundaries between which the charged particle is trapped around the BH and forms a toroidal region. The latest type is characterized by not having any inner and outer boundaries, where the particle has the possibility of being trapped by the BH or escaping into infinity. The impact of parameter $\alpha$ on particle trajectories can explicitly be observed in Fig. \ref{Traj3}. Also, one can see from the first and second rows of Fig.\ref{Traj2}}, that the MOG field shrinks the region between outer and inner boundaries, and increases the chaos in the bounded orbits. While it is for negatively charged particles with ${\cal B}<0$ vise versa. However, the wideness of the bounded orbits' region for negatively charged particles is smaller than it is for positively charged particles.  
\section{Particle acceleration in the magnetic Penrose process}\label{sec4}
Particles within a neutral accretion disk can be ionized in a variety of ways, such as particle disintegration or atom ionization due to particle collision in the hot and magnetized accretion disc. Another alternative possibility is connected with the accretion disk containing a quasineutral collection of ions, and electrons as charged particles in circle-like orbits around BH as well as photons. When the matter in the disk is sufficiently thick, the charged particles' principal free path becomes significantly shorter compared to the entire circumference of orbits, and they travel together like a neutral entity. The plasma density decreases dramatically along the disk near the inner boundary, and charged particles are not further restricted by the surroundings and start floating independently, controlled entirely via the (electro)magnetic fields.
	
This particle ionization model, proposed by Stuchl{\'\i}k et al. in \cite{Stu-Kol:2016:EPJC:}, corresponds naturally to the magnetic Penrose process (MPP) \cite{Par-Wag-Dad:APJ:1986:,Dad-etal:2018:MNRAS:}, in which the initial neutral particle is divided into two charged particles. This straightforward ionization model may be used to investigate the destiny of ionized Keplerian disks \cite{Stu-Kol-Tur:2019:MDPI:}, and if the magnetic influence is weak, it can be used to create modest oscillations of particles in circular orbits \cite{Tur-Stu-Kol:2016:PHYSR4:,Kol-Tur-Stu:2017:EPJC:}. The ionization process conserves both the charge and kinetic momentum of the particle \cite{Stu-etal:2020:Universe:}
\beq
0 = q_2 + q_3 , \quad \cp_{\alpha(1)} = \cp_{\alpha(2)} + \cp_{\alpha(3)}.
	\label{MPP}
\eeq

It is possible to demonstrate that kinetic momentum remains unchanged during ionization
\beq
p_{\alpha(1)} = p_{\alpha(2)} + p_{\alpha(3)},  \label{MPPa}
\eeq
because the electromagnetic effects balance out \cite{Stu-Kol:2016:EPJC:,Stu-Kol-Tur:2019:MDPI:}. In practical cases, one of the generated charged particles might be significantly heavier than the other; for example, in the ionization of atoms, the ion is much more massive than the electron. The gradually more massive charged product absorbs practically all the kinetic momentum of its predecessor's neutral particle, and the lower-charged product's dynamic impact may be ignored.
\beq
p_{\alpha(1)} \approx p_{\alpha(2)} \gg p_{\alpha(3)}. \label{IonMech}
\eeq
The neutral particle will be separated into two charged products after ionization, and while their total mechanical momentum remains conserved, the charged particle will feel the effect of the electromagnetic field via Lorentz force, and their trajectories will be pretty different from the neutral one, as shown in Fig. \ref{MPP1}.
	
In this section, we assume that the ionization event would take place close to the Kerr-BH equatorial plane but with some small perturbation $\theta_{0}\simeq\pi/2$, which allowed the particle to move also in the vertical direction. The neutral parent particles appear to be in a spherical orbit having a starting location $x^\alpha$ and four-velocity $u_\alpha$,
\begin{eqnarray}
x^\alpha&=&(t,r,\theta,\phi)=(0,r_0,\theta_0,0),\\
u_\alpha&=&(u_t,u_r,u_\theta,u_\phi)=(\ce,0,0,\cl).
\end{eqnarray}

The particular energy $\ce$ and angular momentum $\cl$ associated with neutral electrical particles adhering to circular orbits regulate the particle's motion constants \cite{Kolos:2020ykz}
%
\begin{eqnarray} \label{sphE}
    {\cal E}&=&\frac{r^{2} -2 (1+\alpha) r + \alpha \pm a \sqrt{(1+\alpha) r-\alpha} }{r {\cal P}} ,  
\\ \nonumber \label{sphL} 
 {\cal L} &=&\pm \frac{1}{r {\cal P}}\Big[\pm a \alpha+\sqrt{(1+\alpha) r-\alpha} \\ &\times&\Big(r^2 + a^2 \pm 2 a \sqrt{(1+\alpha) r-\alpha}\Big)  \Big]
  \end{eqnarray}

 where ${\cal P}^2=r^2 -3(1+\alpha) r+2\alpha \pm 2a \sqrt{(1+\alpha) r-\alpha}$.
The inner edge of the Keplerian disks is situated at the ISCO, which can be obtained with the condition of $d^2 \Uf_{eff}/dr^2=0$.
\begin{figure*}\vspace{-0.6cm}
	\includegraphics[width=\textwidth]{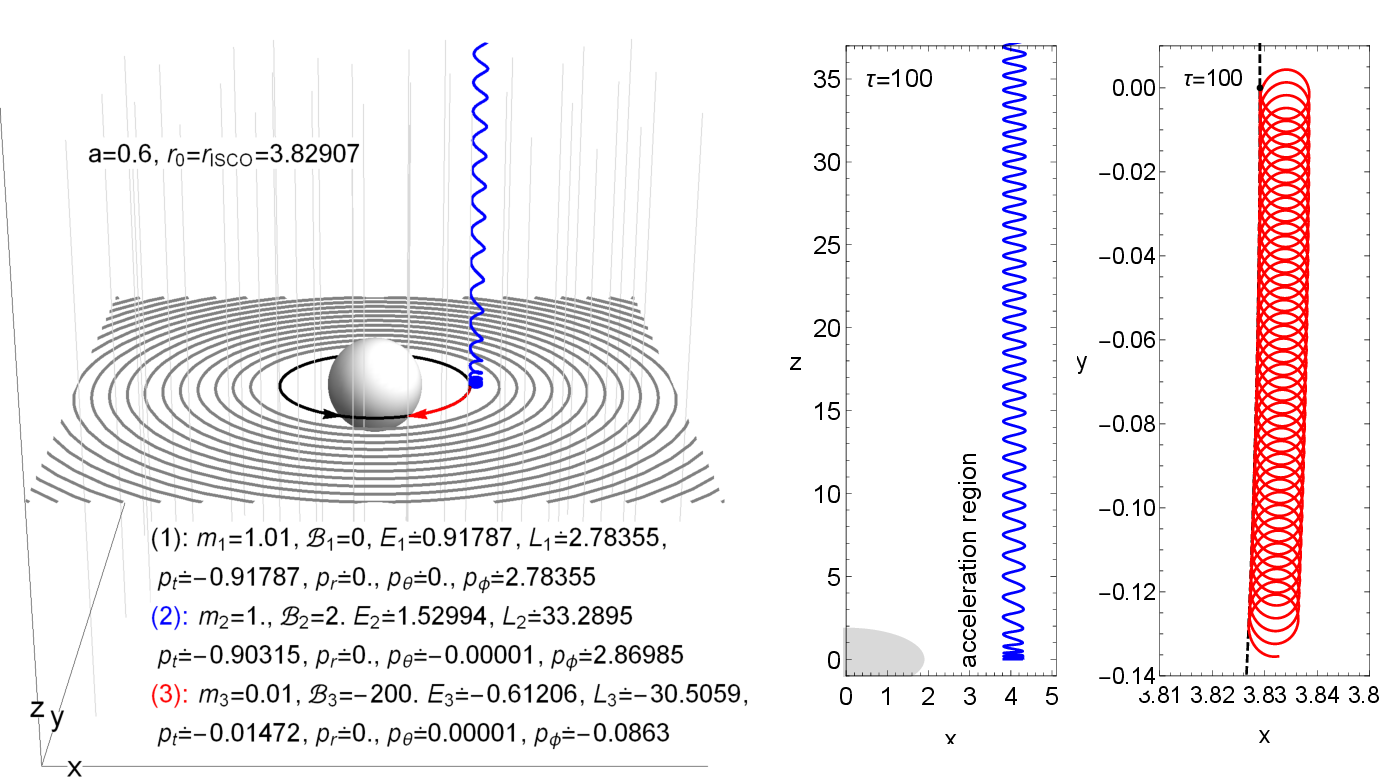}\vspace{0.3cm}
	\includegraphics[width=\textwidth]{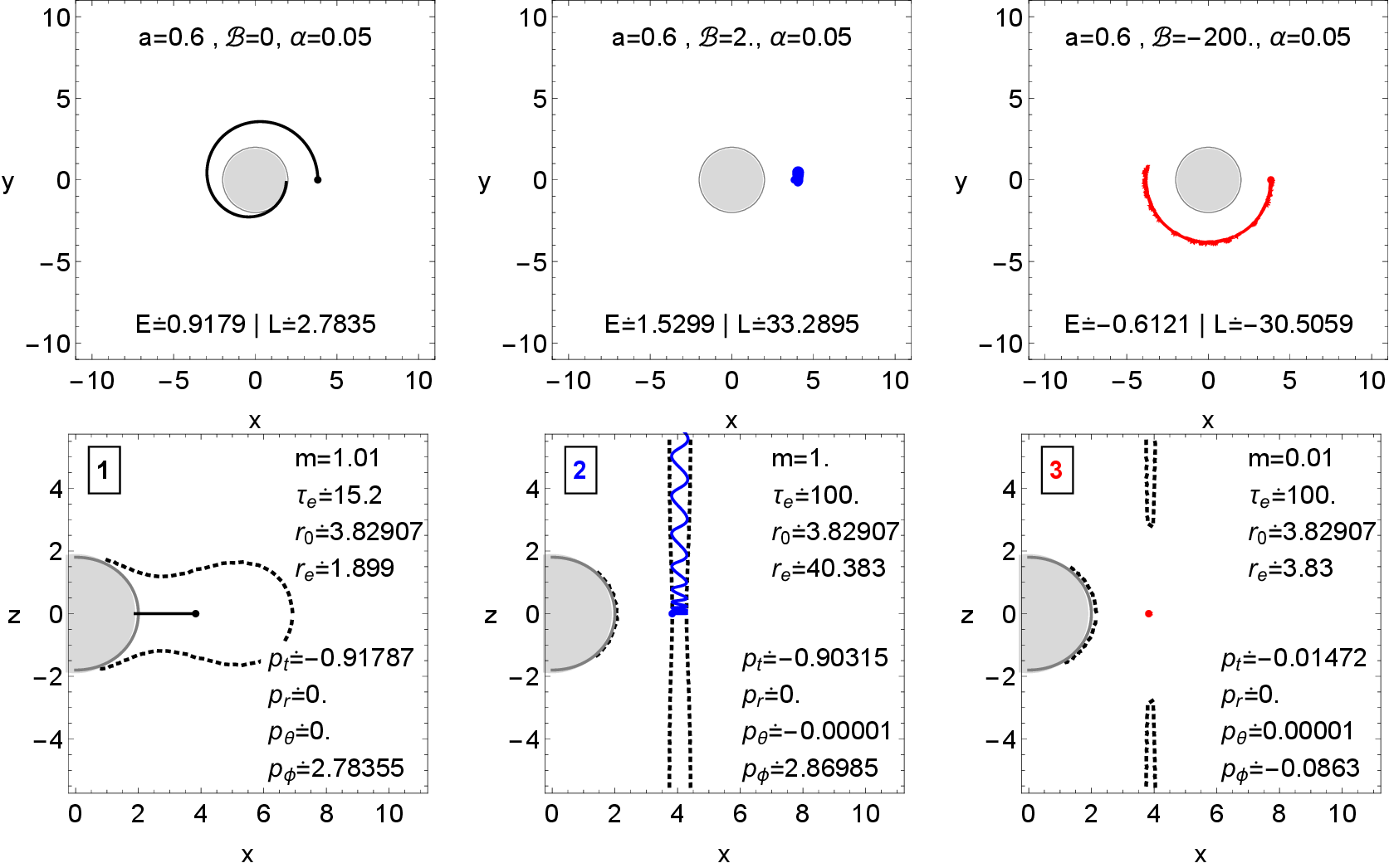}
\caption{Particle acceleration near a magnetized spinning BH.  In the left column, the electrically neutral particle (or in the absence of the external magnetic fields) is shown in solid black curves, which lie along the inside edge of the accretion disk, decomposing into two charged particles, in the middle column positively charged (blue curves) and in the right one negatively charge particles (red curves). Whereas the third particle (red) gets stuck with an excessive amount of negative energy upon orbiting around BH, the second particle (blue) gains an excessive amount of energy via chaotic scattering and escapes along the magnetic field line.}\label{MPP1}
\end{figure*}
\begin{figure*}
	\includegraphics[width=\textwidth]{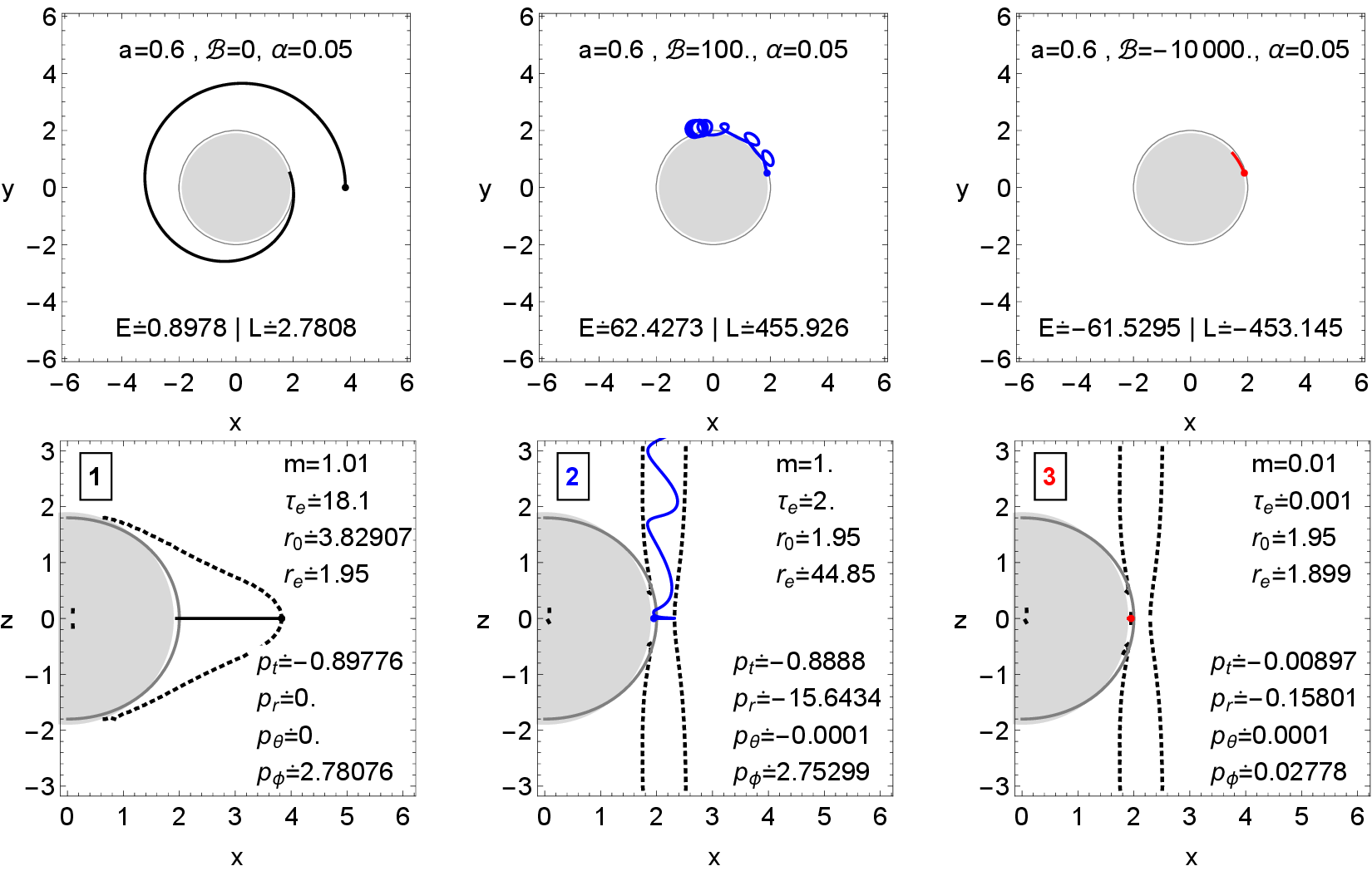}
	\caption{Identical scenario to that of Figure \ref{MPP1}, however this time with comparatively higher values of the magnetic field parameter $\BB$.}\label{MPP2}
\end{figure*}

\subsection{Magnetic Penrose process}

Although the MPP is a regional decaying procedure, consequently, its energy balance can only be determined with the help of the electromagnetic field's local value. As a result, a simple estimation of an asymptotically homogeneous magnetic field compatible with a spinning axis could be utilized. The neutral particle (neutron) orbiting central BH in a thin Keplerian disk or thick torus is in a bound state and its energy is slightly lower than one $E_1<1$, the value $E_1=1$ reserved for a particle at rest located at infinity. Thus, the second particle's (proton) energy, $E_2 = p_{t2} + qA_t$, has the potential to be quite high, while the third particle's energy, $E_3 = p_{t3} - qA_t$, could be negative and of extremely huge magnitude. MPP combined with chaotic dynamics in the combined gravitational and magnetic field leads to the second charged particle trajectory \cite{Stu-Kol:2016:EPJC:} and the circular motion of the particles will be transmuted into linear motion alongside the lines of the magnetic field. Such a flow of charging particles may be used as a straightforward model for the relativistic jets (or particle winds) that have been discovered in several active galactic nuclei and quasars. The BH instantly captures the 3rd particle that has a tremendous negative energy, as illustrated in Fig. \ref{MPP1}.
	
The MPP of the charged particles that accelerate is depicted in Figs. \ref{MPP1} and \ref{MPP2}, at various values of $\cb$.
As previously discussed in \cite{Stu-Kol:2016:EPJC:}, the mechanism of chaotic scattering ensures that the extremely high energy 2nd particle is able to depart to infinity through magnetic field routes, while the particle with significant negative energy is instantly absorbed by the BH (see Figs. \ref{MPP1} and \ref{MPP2}). Moreover, it should be observed that in case of a greater value of the magnetic field, the chaotic nature increases, and the particle escapes to infinity more quickly.
\subsection{Astrophysical relevance \label{Srelevance} }

In practical astrophysical situations, the arbitrary electromagnetic field near a BH is insufficient because its stress-energy tensor does not contribute to the geometry of spacetime. In order to contribute, the electromagnetic field energy density should be in the order of gravitational field density. Therefore, the requirement for the existence of a magnetic field having strength $B$ and BH mass $M$ is outlined below \cite{Tursunov:2020juz,Tursunov2020ApJ...897...99T}

\beq \label{BBB}B << B_{\rm G}=\frac{c^4}{G^{3/2} M_{\odot}} \left(\frac{{M}_{\odot}}{M}\right)\aprx 10^{19}{ \frac{{M}_{\odot}} {M} }\,{\rm Gauss}\, .\eeq
However, it may have a significant influence on the motion of a charged particle. The interplay of the electromagnetic Lorentz force and gravitational attraction experienced by charged matter symbolized by a specific charge $q/m$ could be expressed using the dimensionless "magnetic parameter" $\cb$.
Charged test particles may symbolize protons, electrons, ions, and massively charged anomalies such as plasma objects or charged dust with specific charges $q/m$ ranging from the electron maximum to zero. Due to the enormous magnitude of its specific charge, the magnetic field parameter $\cb$ for protons, ions, and, in particular, electrons can be enormous even in relatively weak magnetic fields, setting up an essential impact of the electromagnetic Lorentz force on its dynamics even in weak magnetic fields. An analogous study that involves protons and ions is shown in Table~\ref{PPval}.
\begin{table}[h]
\begin{tabular}{c@{\quad} c@{\quad} c@{\quad} c@{\quad} c }
	\hline
	$M=10 M_{\odot}$/ & electron & proton & Fe+ & charged dust \\
		\hline
$\cb = 0.1$ & $10^{-4}~{\rm{G}}$ & $ 0.4~{\rm{G}} $ & $ 24~{\rm{G}}$ &  $ 10^{10}~{\rm{G}}$\\
			\hline
	\end{tabular}
\caption{Magnetic field strength $B$ related to magnetic parameter $\cb=0.1$ for different kinds of charged particles travelling near the BH $M_{\rm BH}\sim{}10~M_{\odot}$. \label{PPval} }
\end{table}

Suppose the MPP is associated with an ionized Keplerian disk and chaotic scattering approach. In that case, one might deduce that magnetized revolving BHs could result in the formation of jets headed towards infinity with exceptionally high velocities as a result of the rotational energy extraction of BHs due to capturing electrons with significant negative energy as a result; this is a fundamental procedure for an extra challenging Blandford-Znajek mechanism \cite{Dad-etal:2018:MNRAS:}.
	
When the ionized disk rotates around a non-spinning BH (or a slowly spinning BH encircled by a relatively weak magnetic field), the MPP creates storms that cannot go on to infinity, with acquired energy from the rotational motion of the encircling objects (which represents the Payne-Blandford procedure \cite{Bla-Pay:1982:MNRAS:}). The chaotic scattering procedure resulted in the creation of jets caused by the conversion of revolving energy of matter initially revolving in the Keplerian disk; nevertheless, the energy associated with this transformation of circular motion energy to translational dynamic energy is smaller than that affiliated with electromagnetic field acceleration.
\subsection{Ultra-High Energy Cosmic Rays in the Extreme Regime as Products of MPP}
Cosmic rays are connected to the significant growth of particles’ energy. These are typically made up of high-energy ions or protons; their detected isotropic dispersion implies an extragalactic origin, which is why the production mechanism has long been contested. Inspections of Ultra-High-Energy Cosmic Rays (UHECRs) associated with energetic particles of $E > 10^{18}$ eV—and, on rare occasions, particles with energy $E > 10^{21}$ eV are found, surpassing the GZK limit of $10^{19}$ eV owing to the interplay of cosmic microwaves background—are of particular interest \cite{Stu-etal:2020:Universe:}.
The study of particles having larger energy than the GZK limit necessitates strict distance limitations on the origin of such energetic particles.

The extreme acceleration of particles with energies $E > 10^{21}$ eV can be hard to explain. However, we may suggest a basic strategy centered on the MPP's ultra-efficient regime if it acts in proximity to a supermassive BH surrounded by an abundant magnetic field. The charged particle's energy generated by the MPP in an extraordinarily efficient regime may be described as
\begin{eqnarray}
\nonumber
    E_{MPP} &=& 1.3 \times 10^{21} eV 
    \\ &\times& \left(\frac{m_p}{m}\right) \left( \frac{q}{e} \right)\left(\frac{M}{10^{10} M_{\odot}}\right) \left(\frac{a B}{10^4 G} \right).
\end{eqnarray}

In the preceding formula, $m$ and $q$ stand for the mass and charge of the test particle, respectively, whereas $e$ and $m_p$ symbolize the charge and mass of a proton. Protons possessing an energy $E > 10^{21}$ eV are undoubtedly possible even when $a=0.8$, supermassive BHs possess mass $M = 10^{10} M_{\odot}$ is contained in a magnetic field with strength $B = 10^4 G$ and is quietly rotating.
It is important to note that the energy of a photon may be calculated with respect to the supermassive BH SgrA$^*$, and the corresponding magnetic field observable in the Galaxy's center is provided by
\begin{eqnarray}
    E_{p}^{\text{SgrA}^*}&=&10^{15.6} eV \\ \nonumber &\times& \left( \frac{q}{e} \right) \left( \frac{a}{0.4} \right)
\left(\frac{B}{10^{10} G} \right) \left(\frac{M}{4\times 10^{6} M_{\odot}}\right) \left(  \frac{m_p}{m} \right).
\end{eqnarray}

It is really attractive that this quantity is comparable to the well-known knee of the energy range in the gathered data, which is situated at $E_{knee} \sim 10^{15.6}$ eV, in which the entire amount of particle flow detected is greatly reduced, implying the presence of an incredibly strong only origin placed at a relatively close distance. The MPP on SgrA$^*$ recommends that the model be connected to the UHECR data knee at $E \sim 10^{15.6}$ eV.
	\section{Concluding remarks}\label{Concluding}
The magnetic field has a profound influence on astrophysical phenomena that occur in the vicinity of BHs and other compact objects. Even a tiny magnetic field may considerably affect the location of an ionized Keplerian accretion disk's inner edge and the charged particle's trajectory if the particular particle's charge $q/m$ is sufficiently high. As GRMHD algorithms reveal, the real magnetic field near a BH may possess a relatively complex nature \cite{Komissarov1,Komissarov2}. Therefore, this article examines charged-particle motion and acceleration in MPP around a Kerr-MOG BH drowned into a uniform magnetic field configuration.

From the investigations of horizons, we observed that Kerr BH has the smallest, whereas Schwarzschild-MOG BH has the most enormous horizons. The graphs also reflected that $\alpha$ contributes to the area of BH horizons, whereas $a$ results in shrinking it.
To have information on the circular orbit's stability, we have explored the effective potential. Our study showed that the incoming particle from infinity with a larger value of $a$ required more energy to climb the effective potential in comparison with its smaller value and vice versa for the parameter $\alpha$.
The second and third panels of Fig. \ref{Hor} demonstrate that the circular orbits are initially unstable but become stable along the axes. We also observed that $\alpha$ results in decreasing circular orbit instability near the horizon of BH. The instability of circular orbits near Kerr-MOG BH's horizons is higher than the Schwarzschild-MOG BH. Interestingly, the magnetic field contributes to the stability of circular orbits.

It has been observed that particles have structural trajectories on the equatorial plane, while their nature becomes chaotic as the angle of inclination varies from the equatorial plane. In principle, charged particle dynamics around a magnetized BH reveal four feasible regimes of the ionized Keplerian disk behavior, i.e., surviving in consistent epicyclic motion, changing into a chaotic toroidal state, collapsing due to escaping along the magnetic field paths and collapsing due to falling into the BHs.
From the study of particle trajectories, we noticed that at fixed values of other parameters, Schwarzschild BH captures the test particle, whereas, in Kerr BH, the test particles may either flee to infinity or be trapped by the BH, while in Kerr-MOG BH, the test particle is trapped in some region around BH and starts orbiting it (see Fig. \ref{Traj1}).

Our study shows various types of boundary conditions. The first one is related to the existence of an outer boundary, where the particle should be captured by the BH. The second type is related to the existence of an inner boundary where the particle must escape to infinity. The 3rd type is related to the existence of inner and outer boundaries between which the charged particle is trapped around the BH and forms a toroidal region. The last type is characterized by not having any inner and outer boundaries, where the particle has the possibility of being trapped by the BH or escaping into infinity. The impact of parameter $\alpha$ on particle trajectories can explicitly be observed in Fig. \ref{Traj3}.

We investigated an approach for supermassive BHs to be responsible for the production of UHECRs. Using a unique, ultra-efficient regime of MPP and ionization of neutral particles, including neutron-beta decaying at the horizon of a rotating BH. On investigating the MPP, we found that for a larger magnetic field, the behavior of orbits becomes more chaotic due to the particle escaping towards infinity more quickly, which is analogous to the findings of \cite{Tursunov:2020juz}.
Besides, charged particles from ionized Keplerian disk may be accelerated near Kerr-MOG BHs with an MPP efficiency of more than $10^{10}$, allowing protons to be accelerated to energy:\\
$\bullet$ $10^{21}$ eV in the vicinity of a supermassive BH of mass $M = 10^{10}M_{\odot}$ and $B = 10^{4}$G;\\
$\bullet$ $10^{19}$ eV in the vicinity of M$87$, a supermassive BH of mass $M = 7\times 10^{9}M_{\odot}$ and $B=10^{2}$ G; and \\
$\bullet$ $10^{15.6}$ eV in the vicinity of SgrA$^*$, a supermassive BH of mass  $M = 4\times 10^{6}M_{\odot}$ and $B=10$ G.\\
This approach may be used for neutron stars with small masses compensated by enormous magnetic fields. However, such types of research will be left for future research. Because relativistic electrons lose synchrotron radiation $10^{10}$ times quicker than protons, heavier elements of UHECRs appear more likely in this scenario.

To predict the candidates of supermassive BHs within the proposed framework, magnetic field data at the event horizon scale is required. There are currently few precise measurements, but future worldwide VLBI observations will likely raise this number. We believe that the proposed concept of supermassive BH as a UHECR power engine opens up new avenues to comprehend this unique high-energy phenomenon and its significance in a variety of high-energy scenarios.
\section*{Acknowledgments}
JR acknowledges Grant No. FA-F-2021-510 of the Uzbekistan Agency for Innovative Development and the ERASMUS+ ICM project for supporting his stay at the Silesian University in Opava.
	\subsection*{Conflicts of Interest}
The authors declare no conflict of interest.
	
	\def\prc{Phys. Rev. C}
	\def\pre{Phys. Rev. E}
	\def\prd{Phys. Rev. D}
	\def\prl{Physical Review Letters}
	\def\jcap{Journal of Cosmology and Astroparticle Physics}
	\def\apss{Astrophysics and Space Science}
	\def\mnras{Monthly Notices of the Royal Astronomical Society}
	\def\apj{The Astrophysical Journal}
	\def\aap{Astronomy and Astrophysics}
	\def\actaa{Acta Astronomica}
	\def\pasj{Publications of the Astronomical Society of Japan}
	\def\apjl{Astrophysical Journal Letters}
	\def\pasa{Publications Astronomical Society of Australia}
	\def\nat{Nature}
	\def\physrep{Physics Reports}
	\def\araa{Annual Review of Astronomy and Astrophysics}
	\def\apjs{The Astrophysical Journal Supplement}
	\def\aapr{The Astronomy and Astrophysics Review}
	\def\procspie{Proceedings of the SPIE}

	\bibliographystyle{spphys}
	\bibliography{reference}
\end{document}